\documentclass[preprint,12pt]{elsarticle}

\usepackage{graphicx,epsfig,epsf,amssymb,natbib}
\usepackage{amssymb}
\usepackage{url}

\begin{document}

\begin{frontmatter}

\title{Comparison of Fermi-LAT and CTA in the region between 10-100 GeV}
 
\author[SLAC]{S. Funk\corref{corauths}}
\author[LE]{J.~A. Hinton}
\author[CTA]{for the CTA Collaboration}

\address[SLAC]{Kavli Institute for Particle Astrophysics and
  Cosmology, Department of Physics and SLAC National Accelerator
  Laboratory, Stanford University, CA-94305, USA}
\address[LE]{Department of Physics and Astronomy, University of
  Leicester, Leicester LE1 7RH, UK}

\cortext[corauths]{Corresponding author: Stefan.Funk@slac.stanford.edu}

\begin{abstract}
  The past decade has seen a dramatic improvement in the quality of
  data available at both high (HE: 100 MeV to 100 GeV) and very high
  (VHE: 100 GeV to 100 TeV) gamma-ray energies. With three years of
  data from the Fermi Large Area Telescope (LAT) and deep pointed
  observations with arrays of Cherenkov telescope, continuous spectral
  coverage from 100 MeV to $\sim10$ TeV exists for the first time for
  the brightest gamma-ray sources. The Fermi-LAT is likely to continue
  for several years, resulting in significant improvements in high
  energy sensitivity. On the same timescale, the Cherenkov Telescope
  Array (CTA) will be constructed providing unprecedented VHE
  capabilities. The optimisation of CTA must take into account
  competition and complementarity with Fermi, in particularly in the
  overlapping energy range 10$-$100 GeV. Here we compare the
  performance of Fermi-LAT and the current baseline CTA design for
  steady and transient, point-like and extended sources.
\end{abstract}

\begin{keyword}
  Gamma rays: general
\end{keyword}

\end{frontmatter}

\section{Introduction}
\label{intro}

The energy range between 10~GeV and 100~GeV is the range where
space-based satellites such as the Fermi-LAT ($\sim20$~MeV to
$\sim300$~GeV) and ground-based instruments such as the
next-generation Cherenkov Telescope Array (CTA) overlap. While they
have vastly different effective areas, their sensitivity in this
overlapping region is similar (differential flux $\sim10^{-12}$ erg
cm$^{-2}$ s$^{-1}$ for a typical 100 hour observation and a 10-year
Fermi-LAT mission) since the Fermi-LAT is close to background-free at
those energies. This energy range is of great importance for a wide
range of scientific topics:
\begin{itemize}
\item the Universe goes from being transparent to gamma-rays to having
  a pronounced horizon at a redshift $<1$~\cite{Gould1966,
    Jelley1966, EBL:FazioStecker}, limiting the number of bright very
  distant objects (such as e.g.\ Gamma Ray Bursts, or GRBs) that can
  be studied, but providing information on the cosmological evolution
  of the infrared-ultraviolet background light~\cite{HESSEBL1,
    HESSEBL2, FermiEBL, Aharonian2001, Primack2001, Dwek2001}.
\item the brightest Fermi-LAT sources~\cite{Fermi1FGL} have spectra
  that steepen in this energy range. Examples for this behavior are
  flat-spectrum radio quasars~\cite{Fermi1LAC},
  and mid-aged supernova remnants
  interacting with molecular clouds~\cite{FermiW51C, FermiW44}
\item the diffuse Galactic background has a steeper spectrum than most
  detected Fermi-LAT sources and therefore goes from being the
  dominant $\gamma$-ray emitter below this energy range (at $\sim
  100$MeV) to being sub-dominant to individual sources in the TeV
  band~\cite{HESS:scanpaper1, HESS:scanpaper2, HESS:gc_diffuse,
    Fermi:diffuse, milagro:diffuse}.
\item the spectra of a large number of Galactic object source classes
  exhibits rising components (in energy flux) and the emergence of
  additional components. Examples for such spectral properties are LS
  5039~\cite{Fermi:LS5039}, Eta Carinae~\cite{Fermi:EtaCar,
    Walter:EtaCar}, the Supernova remnant
  RX\,J1713.7$-$3946~\cite{Fermi:1713} or the Pulsar Wind Nebula
  HESS\,J1825$-$137~\cite{HESS:1825p2, Fermi:1825}. For this reason
  simple extrapolations of the energy spectra into this region from
  above or from below are often expected to be wrong (see
  e.g.~\cite{Funk2008} for a more detailed study of this topic in the
  pre-Fermi-LAT era).
\end{itemize}
In addition, to these scientific topics, cross-calibration between the
Fermi-LAT and ground-based instruments might be
interesting~\citep{CrossCalibration, CrossCalibrationCrab}. The
increased overlap in energy coverage between CTA and Fermi-LAT
compared to current instruments might render this important and might
provide an cross-calibration for ground-based instruments.

\begin{figure}[h]
\mbox{\epsfig{file=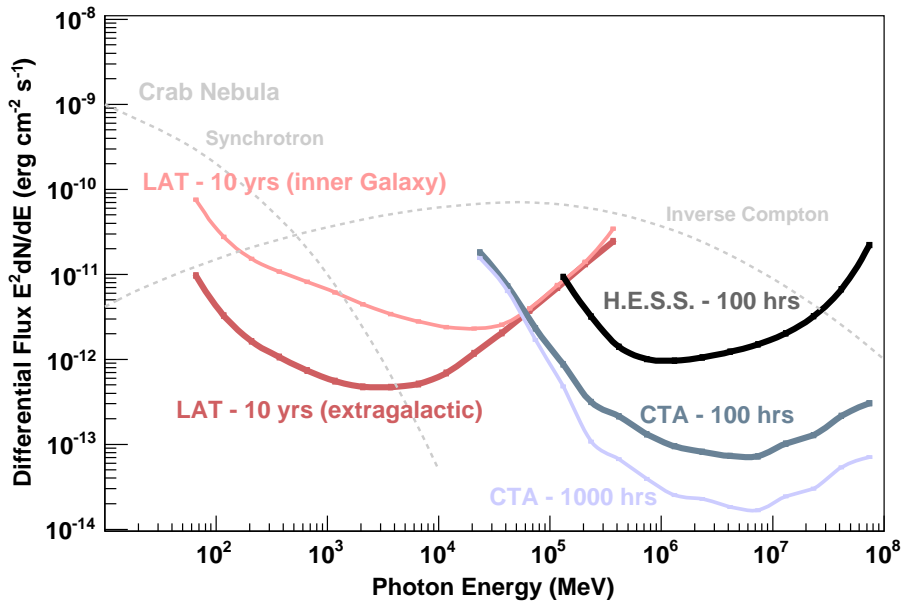, width=0.9\textwidth }}
\caption{``Differential'' sensitivity (integral sensitivity in small
  energy bins) for a minimum significance of $5\sigma$ in each bin,
  minimum 10 events per bin and 4 bins per decade in energy. For
  Fermi-LAT, the curve labeled ``inner Galaxy'' corresponds to the
  background estimated at a position of $l = 10^{\circ}, b =
  0^{\circ}$, while the curve labeled ``extragalactic'' is calculated
  using the isotropic extragalactic diffuse emission only. For the
  ground-based instruments a 5\% systematic error on the background
  estimate has been assumed. All curves have been derived using the
  sensitivity model described in section~\ref{sec:model}. For the
  Fermi-LAT, the {\emph{pass6v3}} instrument response function curves
  have been used. As comparison, the synchrotron and Inverse Compton
  measurements for the brightest persistent TeV source, the Crab
  Nebula are shown as dashed grey curves.
\label{fig:1}
}
\end{figure}

Figure~\ref{fig:1} shows the sensitivity curve of the Fermi-LAT,
H.E.S.S., and CTA for an $E^{-2}$ type power-law spectrum of a source
for several different observation lengths. Here and in the following
the altitude of CTA is assumed to be 2000m. Different altitudes will
change the CTA sensitivity curves slightly but we do not expect the
results described here to change in any significant way. The exact
details of the sensitivity for CTA in general depend on the as of yet
unknown parameters like the array layout and analysis technique of
CTA. However, we don't expect the sensitivity of CTA or the lifetime
of the Fermi-LAT to change by a significant factor compared to what is
assumed here (unless there is a significant increase in the number of
telescopes for CTA).  As the differential sensitivity curves for these
instruments are usually only provided for 1-year of Fermi-LAT and for
50~hours of H.E.S.S./CTA, we had to make use of a sensitivity model
which will be described in section~\ref{sec:model}.  Generally, the
sensitivity information provided is insufficient to make a detailed
comparison of the performance in the overlapping region which
motivates this study.

As can be seen from the figure, the Fermi-LAT is photon starved in the
overlapping energy range and therefore the $\nu F_{\nu}$ (which is
equivalent to $E^2dN/dE$) sensitivity worsens with increasing energy
proportional to $E^{1}$. The Fermi-LAT 10-year sensitivity is
extremely uneven across the sky, due to the bright diffuse gamma-ray
emission from cosmic-ray interactions in our Galaxy in that energy
range~\cite{Fermi:diffuse}. We show two positions, one labeled "inner
Galaxy" at $l=10^{\circ}, b=0^{\circ}$ Galactic coordinates and one at
high latitudes labeled ``extragalactic'', taking into account only the
isotropic diffuse emission~\cite{Fermi:isotropic}. The Galactic
diffuse emission has a steeper spectrum than $E^{-2}$ and is therefore
increasingly less dominant with higher energies in the
Fermi-LAT~\cite{Fermi:diffuse}. For our study we will ignore the
Galactic diffuse background in the following. This has negligible
effect on the energy at which the Fermi-LAT and CTA differential
sensitivity curves overlap as seen in Figure~\ref{fig:1}. It should be
noted that in the very inner parts of the Galaxy diffuse emission can
become an issue, even for CTA as shown
in~\cite{HESS:gc_diffuse}. Contrary to the Fermi-LAT, CTA is
systematic error dominated in the overlapping energy range. Therefore
longer observations do not help the CTA sensitivity in this range as
can be seen from Figure~\ref{fig:1}. Unless otherwise noted, we have
assumed that the source counts need to be at least 5\% above the
background to be significantly detected (i.e.\ we assumed that we can
determine our background level to 5\% accuracy). While this is a
reasonable assumption, for special observations, such as for pulsars
(where the background can be determined by the off-phase), this might
be overly conservative. Due to the dominance of systematic errors for
CTA in the overlapping energy range, longer observation times do not
significantly shift the energy at which the Fermi-LAT and CTA
sensitivity curves cross as can be seen in Figure~\ref{fig:1}.

\begin{figure}[h]
 \mbox{\epsfig{file=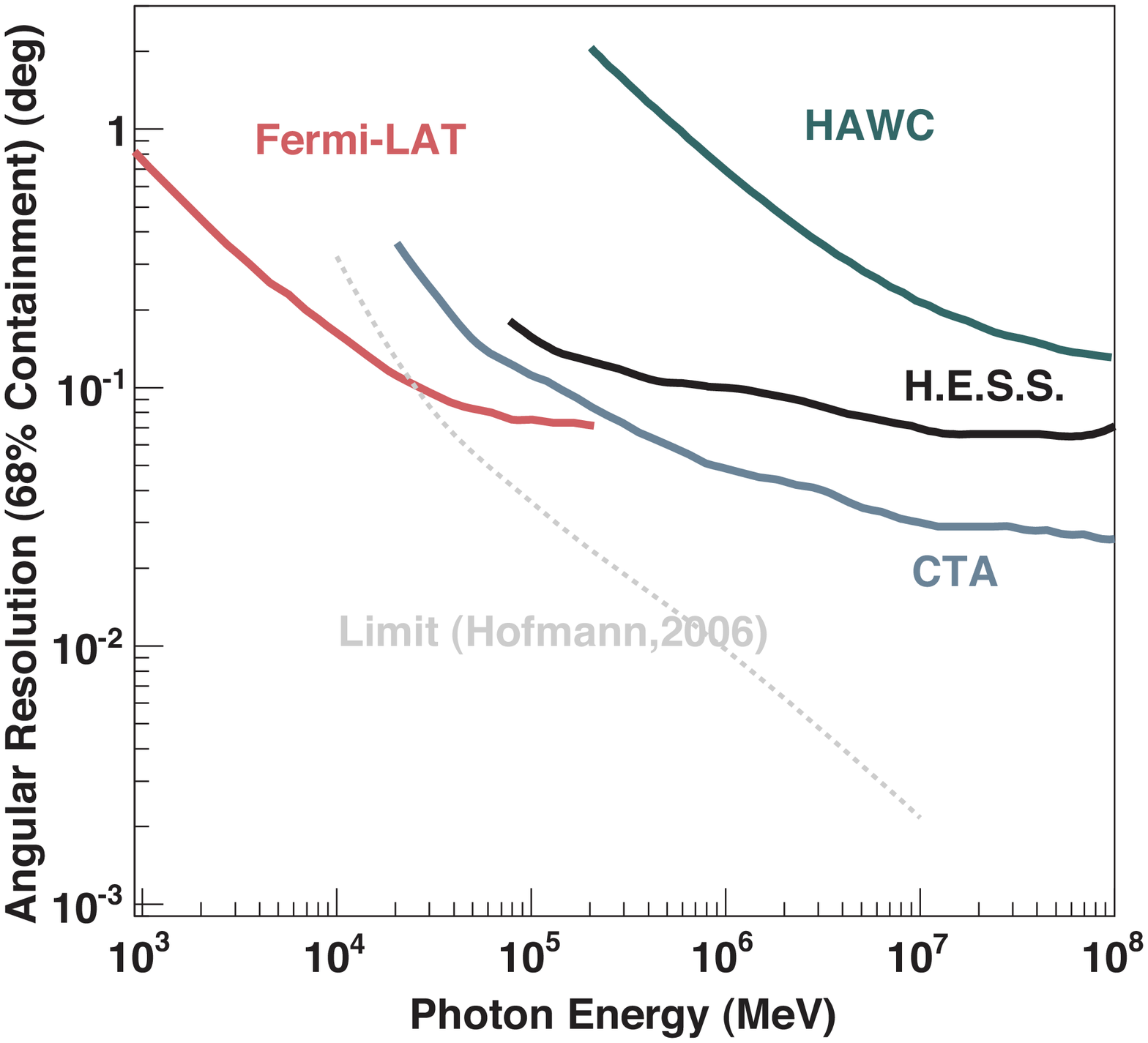, width=0.5\textwidth}}
 \mbox{\epsfig{file=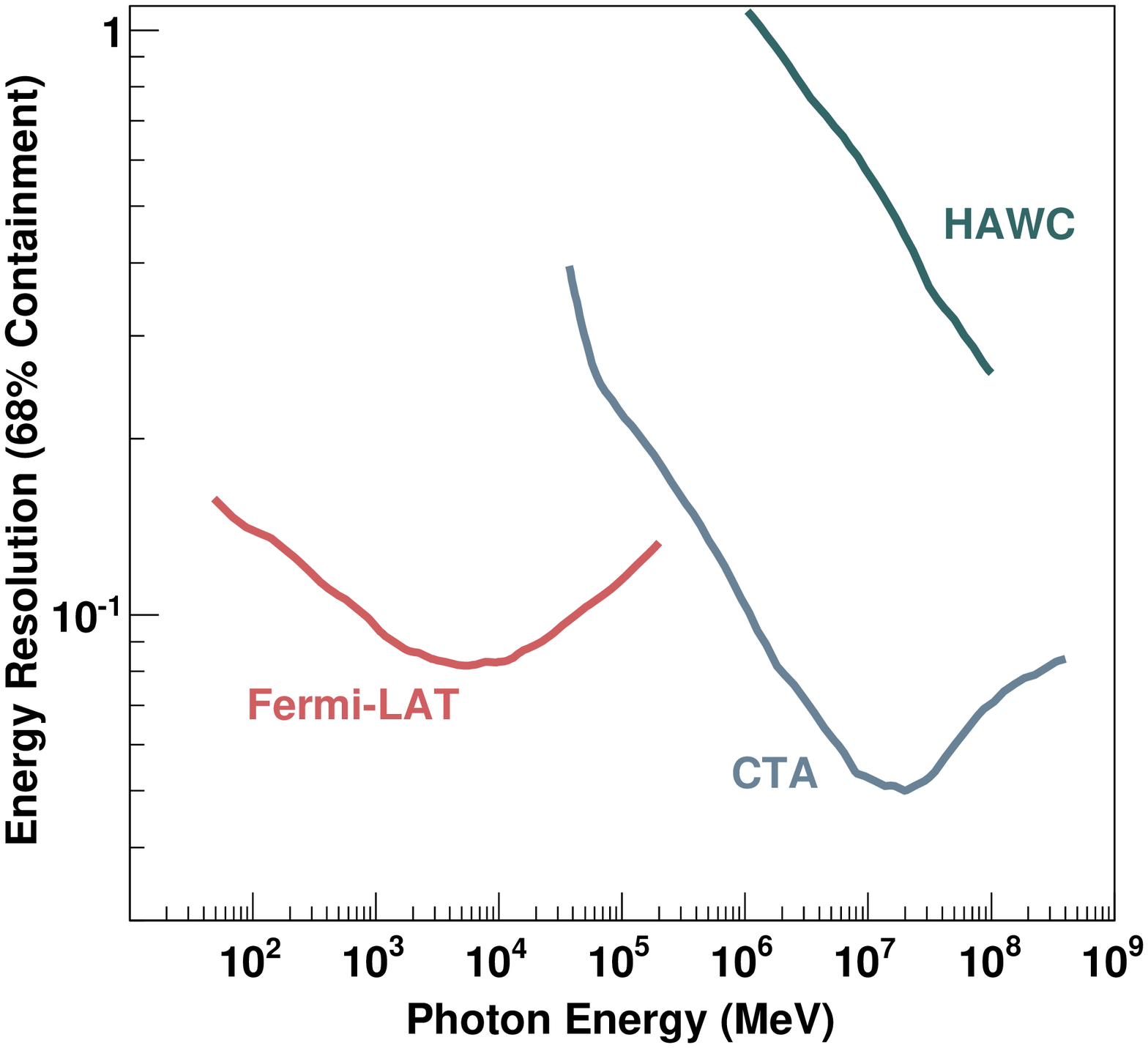, width=0.5\textwidth }} 
 \caption{{\bf{Left:}} Angular resolution for
   Fermi-LAT~\cite{Fermi:performance} and
   CTA~\cite{design_study}. H.E.S.S.~\cite{FunkThesis} and
   HAWC~\cite{HAWC:Performance} are shown as examples for a
   current-generation IACT and for a next-generation water Cherenkov
   detector. Also shown is the limiting angular resolution that could
   be achieved if all Cherenkov photons emitted by the particle shower
   could be detected~\cite{Hofmann:Performance}. The CTA curve has not
   been optimized for angular resolution and enhanced analysis
   techniques are expected to improve this curve. {\bf{Right:}} Energy
   resolution for Fermi-LAT and CTA. Shown is the 68\% containment
   radius around the mean of the reconstructed energy. It is evident
   that the energy resolution of Fermi-LAT in the overlapping energy
   range is significantly better than the CTA resolution. }
\label{fig:2}
\end{figure}

Differential sensitivity is clearly not the only relevant factor when
comparing instruments in the overlapping range. The integral
sensitivity is relevant when aiming to detect a new source, and the
angular and energy resolution are clearly critical for imaging and
spectroscopy. Figure~\ref{fig:2} shows the angular resolution and the
energy resolution for the instruments operating (or planned) in the
$\sim100$ GeV range. As can be seen there are orders of magnitudes
differences between instruments in both quantities. Below 100 GeV the
Fermi-LAT outperforms all ground-based instruments in both angular and
energy resolution. This is due to inherent fluctuations in those
particles above the Cherenkov threshold high in the atmosphere for
showers initiated by low energy primaries. So even if the differential
sensitivity of the Fermi-LAT and CTA is the same at a given energy,
the Fermi-LAT will be able to do a better measurement of a
source. While HAWC's performance in these quantities is rather modest,
its main goal is to detect new sources and study variability and find
transients. HAWC is not shown in Figure 1 as differential sensitivity
curves has not been provided by the HAWC collaboration and indeed, it
is not the relevant quantity for the aforementioned goals. In the
energy range at which this study is focused, HAWC is not competitive
with the Fermi-LAT and CTA except perhaps for the detection of very
short timescale transients such as GRBs.

\section{The SensitivityModel}
\label{sec:model}

The sensitivity of gamma-ray detectors is determined by three basic
characteristics: the effective collection area, residual background
rate and angular resolution, all of which are typically a strong
function of gamma-ray energy. For Fermi-LAT the relevant curves are
taken from \cite{Fermi:performance} for instrument response function
{\emph{pass6\_v3}}, and for CTA from ~\cite{design_study}. It should
be noted that the usage of the enhanced {\emph{pass7}}
response-functions for the Fermi-LAT will not substantially change the
presented results. The difference in effective area above 1 GeV is
$\sim 10\%$.  We also note that the CTA performance is very likely to
improve relative to that shown here, due to analysis improvements and
hardware performance and telescope layout optimization. For a detailed
description of the CTA instrument response function,
see~\cite{MCPaper} in this issue. Detection sensitivity may be limited
by statistical fluctuations of the background, by background
systematics or by the number of detected signal photons.  The
statistical limit is calculated using a maximum likelihood approach,
background systematics in CTA are assumed to have a 1\% rms
\cite{design_study}, and a minimum of 10 photons is always required
for a detection. The instrument point-spread functions (PSFs) are
assumed to be Gaussian for simplicity, with the 68\% containment
radius ($\theta_{68}$) matched to that of the simulated instrument
response. This study builds on that presented in \cite{dm_paper} but
is more precise in that it uses Monte-Carlo estimated background rates
and collection areas for a baseline CTA design ({\emph{layout "E"}})
\cite{design_study} rather than inferred values, derived for an
idealized future Cherenkov array~\cite{kb_97tels}. Array layout
{\emph{E}} is used as an example. This particular configuration uses
three telescope types: four 24 m telescopes with 5$^{\circ}$
field-of-view, 23 telescopes of 12 m diameter with 8$^{\circ}$
field-of-view, and 32 telescopes of 7 m diameter with a 10$^{\circ}$
field-of-view. The telescopes are distributed over $\sim 3$km$^2$ on
the ground. The study presented here uses the curves for an altitude
of 2000m and a zenith angle of 20$^{\circ}$. The residual background
rate adopted for Fermi (unless otherwise stated) is taken
from~\cite{Fermi:performance} and is representative of the isotropic
diffuse emission relevant for high Galactic latitude sources. As
previously stated we ignore the Galactic diffuse emission which is
justified, given its diminishing importance in the Fermi-LAT data
above 10~GeV. The likelihood method adopted is a simplified version of
that used for data analysis: events are binned in energy but counted
(rather than fit) within an energy-dependent aperture. To match the
sensitivity achieved using the standard method a background scaling
factor of 0.6 is applied. This approach is used throughout except for
the case of the source extension studies described in
section~\ref{sec:ext}, where a full treatment is used.

\begin{figure}[h]
\mbox{\epsfig{file=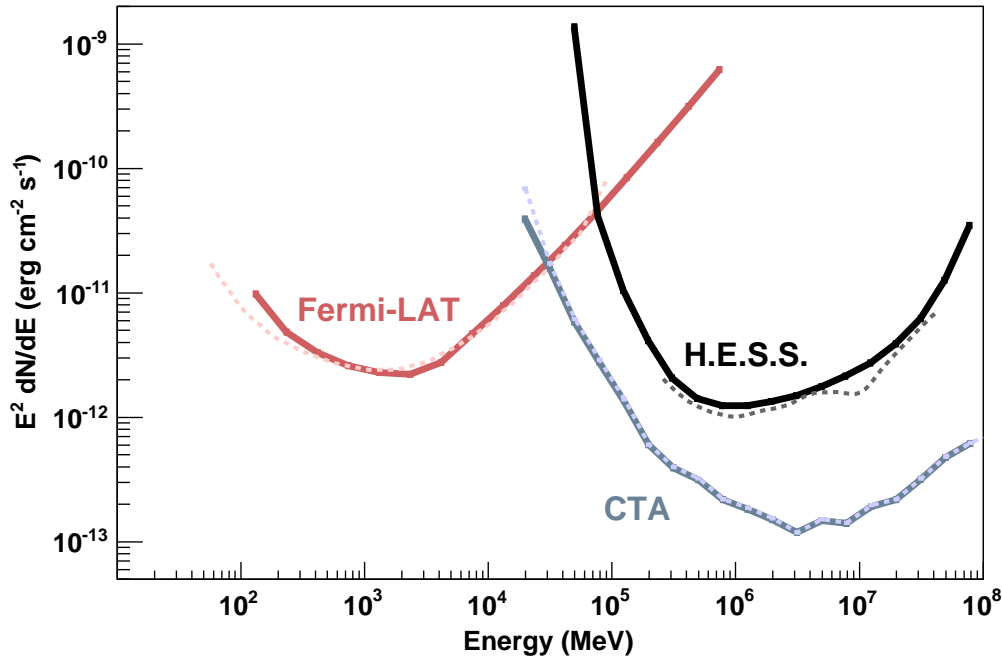, width=1.0\textwidth}}
\caption{Comparison of sensitivities derived with our model to the
  official sensitivity curves available in the
  literature~\cite{Fermi:performance, FunkThesis, design_study}.
  Dotted curves are official curves, solid curves are derived with the
  model described in the text. }
\label{fig:1b}
\end{figure}

In Figure~\ref{fig:1b} we compare the sensitivity model to published
curves for the differential sensitivity of CTA and Fermi, agreement
exists at the 10-20\%, adequate for the purposes of our study, in
particular considering the provisional nature of the CTA curves.


\section{Differential Sensitivity}
\label{differential}

It is interesting to compare differential sensitivities since this is
the relevant quantity when comparing the quality of spectral
measurements, in particular of features such as cutoffs and
breaks. For the differential sensitivities we used a source with
spectral index of $dN/dE \propto E^{-2}$, and required a significance
of 5$\sigma$ in each energy bin. In addition we required the source
flux to be a factor of 5 above the background systematics (which we
assumed to be 1\%). Unless otherwise noted we calculated the
differential sensitivity for 4 bins per decade in energy. In the
energy range under study, both instruments suffer from drawbacks in
spectroscopy: the Fermi-LAT is unable to exploit its good energy
resolution due to a lack of photon statistics, but CTA is unable to
make use of its large collection area due to limited energy
resolution. Given the 30\% energy resolution of CTA, see
Fig.~\ref{fig:2} (right), only 4 independent bins per energy are
possible, assuming separation of the centers of the bins by the
full-width half max of the energy resolution. This should not pose
severe problems, since extremely sharp features are rather rare (apart
from super-exponential cutoffs or dark matter annihilation signals).

When comparing the differential sensitivity in energy bins, clearly
the motivation is to be able to perform spectral measurements. The
energy $E_{\mathrm{cross}}$ at which the differential sensitivity
curves of CTA and Fermi-LAT intersect is the energy below which
Fermi-LAT and above which CTA is better suited to perform spectral
measurements. However, it should be noted that the underlying
motivation is to find the energy at which the Fermi-LAT and CTA
spectral points of a source have similar statistical error bars. For CTA, being
dominated by background systematics in this energy range, a 5$\sigma$
detection requirement in each bin translates (in the Gaussian limit
that applies here) into a 20\% flux error for the point. For the
Fermi-LAT, however, this is not true, since neither the signal nor the
background of a threshold source are in the Gaussian limit in the
energy range under study. Being signal-limited, we estimate that to
get the same flux error, we need $N = 25$ signal events in the energy
bin, significantly larger than what is usually applied when comparing
differential sensitivity curves. If this requirement is fulfilled the
Fermi-LAT should have a comparable error on its spectral measurement.

\begin{figure}[h]
\mbox{\epsfig{file=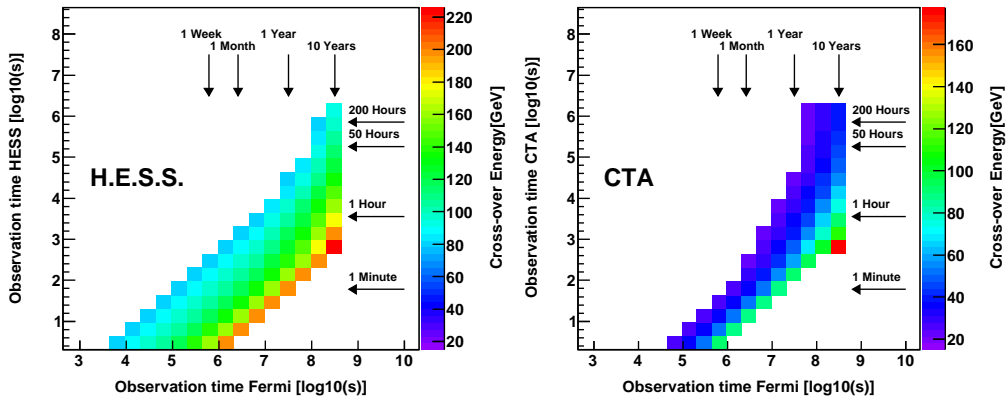, width=1.0\textwidth }}
\caption{Cross over energy $E_{\mathrm{cross}}$ as a function of
  Fermi-LAT and of H.E.S.S. (left) or CTA (right) observation
  time. Here we required a detection significance in each energy bin
  of 5$\sigma$ and a minimum number of 25 events in each bin (to get
  comparable errors on the flux measurement in the bin (see text for a
  discussion). No optimization for loosening the cuts at short
  observation times has been performed (the Fermi-LAT \emph{pass6\_v3}
  diffuse response function was used), therefore in principle the
  Fermi response could be somewhat better - but not by much, given the
  limited physical area of the instrument. Also for CTA, the curves
  could be improved if the systematics are brought under control and
  there was a smaller systematic error on the background estimate.}
\label{fig:3}
\end{figure}

Figure~\ref{fig:3} shows the energy $E_{\mathrm{cross}}$ at which the
error on the flux measurement in the energy bin should be equal
between H.E.S.S. (left) / CTA (right) and the Fermi-LAT as a function
of the observation times in both instruments. Given our assumptions
about the systematic error on the background level, there is typically
no large benefit in the overlapping energy range for CTA to spend
significant amounts of observation time as can be seen from the fact
that $E_{\mathrm{cross}}$ does change very weakly for a given
Fermi-LAT observation time when increasing the CTA observation time.
Also, it can be seen that for an expected 10-year lifetime of the
Fermi-LAT mission $E_{\mathrm{cross}} \sim40$~GeV for the assumed
parameters and for a typical 100~hour CTA observation. That means at
that energy the Fermi-LAT will be doing measurements (within the
10-year mission) that are comparable in quality to the measurements
that will be done in a 100~hour CTA observation.

\begin{figure}[h]
\mbox{\epsfig{file=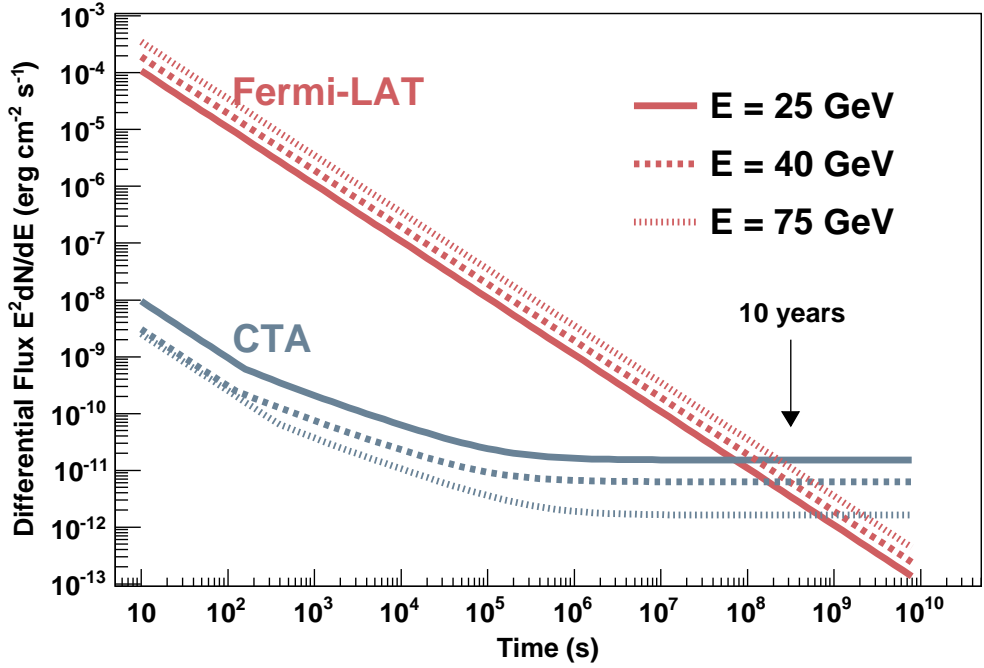, width=1.0\textwidth }}
\caption{Differential sensitivity at selected energies as a function
  of observation time. These plots were generated for a detection
  significance of 5$\sigma$ in the relevant energy bin and a minimum
  number of 25 events.  }
\label{fig:8}
\end{figure}

Clearly, CTA has a huge discovery potential over the Fermi-LAT in the
overlapping energy range for short-transient phenomena (provided they
occur in the field of view), due to the large collection area. To
demonstrate this, we show in Figure~\ref{fig:8} the differential
sensitivity (or more precisely the integral sensitivity in the energy
bin) for selected energies (25, 40, 75 GeV) as a function of
observation time. Since the Fermi-LAT is signal-limited at these
energies, the sensitivity improves rapidly with increasing observation
time. For short-duration transient objects such as GRBs, it is evident
that CTA has an advantage over the Fermi-LAT by many orders of
magnitude which constitutes a large discovery potential. It should
however be said that for transient sources the Fermi-LAT has the
advantage of a 2.4$\pi$ sr field of view which makes catching
transients much more likely. In addition, the Fermi-LAT can view out
gamma rays to much larger redshifts due to the $\gamma\gamma$ pair
production opacity of the Universe at higher gamma-ray energies. The
exact position of the kink in the graphs for CTA depends on the exact
assumptions on the background systematics as discussed above. Again,
it can be expected, that the CTA sensitivity for short-duration events
can be significantly improved compared to the current instrument
response function.

\begin{figure}[h]
\mbox{\epsfig{file=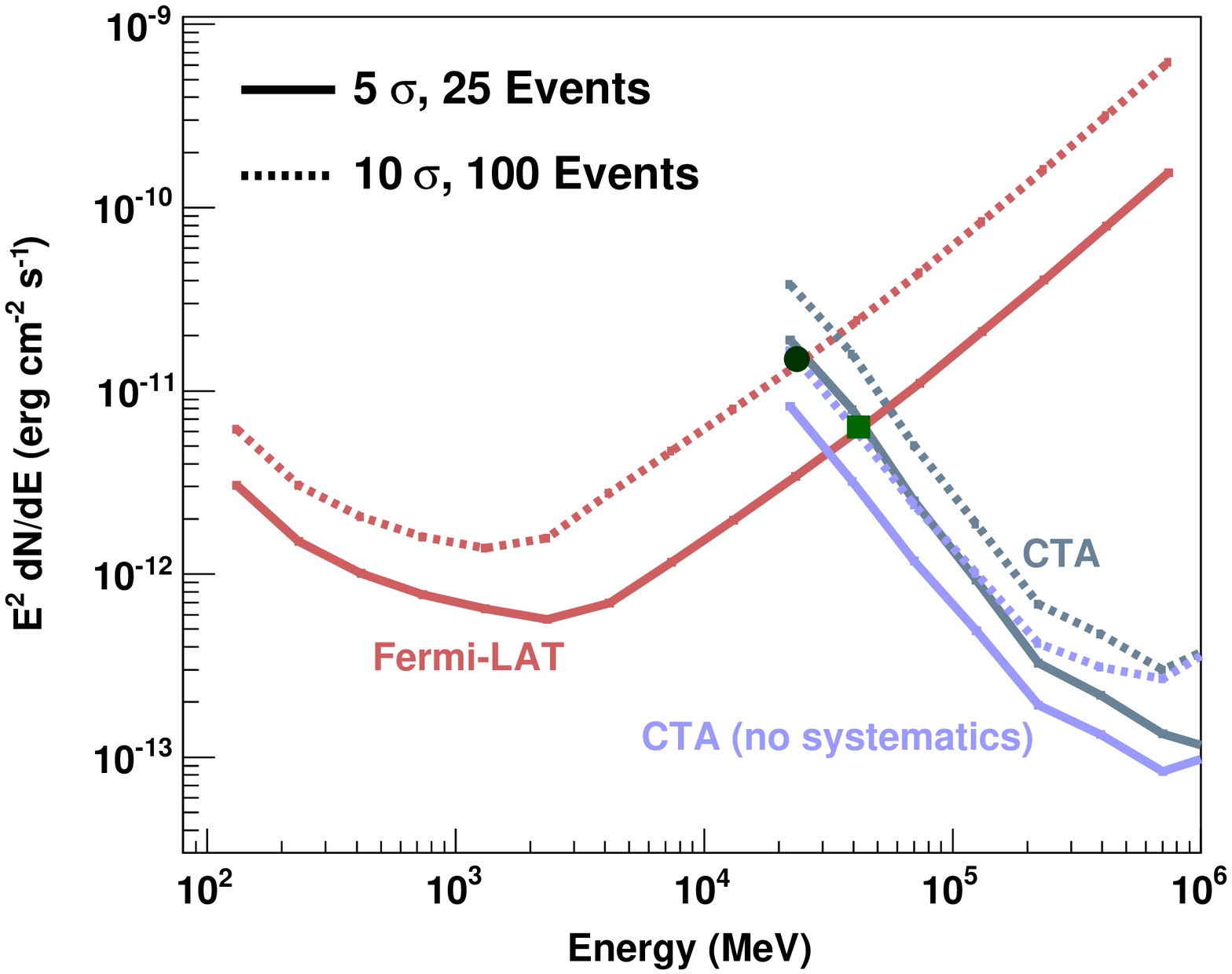, width=1\textwidth }}
\caption{Differential sensitivity for the case of no background
  systematic errors (e.g.\ in pulsar observations) in CTA (light blue)
  and with the standard 1\% systematic error on the background for
  comparison (dark blue). In such cases the aim is to measure the
  spectrum at high precision, therefore 10$\sigma$ per energy bin and
  a minimum of 100 events (resulting in an error on the flux of
  $\sim10\%$) are shown as well (dashed lines). The markers indicate
  the points at which the 10$\sigma$, 100 events, no CTA background
  systematics intersect (dark green circle, 24.2 GeV) and for
  comparison the standard 5$\sigma$, 25 events, 1\% background
  systematics (light green square, 41.8 GeV). The curves are shown for
  10 years of Fermi-LAT and 100 hours of a CTA observation. }
\label{fig:7}
\end{figure}

In addition to faint sources, it is also interesting to consider
bright sources. Recent measurements have established pulsed emission
from the Crab Pulsar in the $>25$GeV range~\cite{MAGICCrab,
  VERITASCrab}.  For such observation the systematic error on the
background level can be significantly reduced, since the local
background can be determined from the off-phase of the pulsar. In this
case the aim is no more the detection in each energy bin, but rather a
very small error on the measured flux. In Figure~\ref{fig:7} we
illustrate the effect of requiring 10$\sigma$ per energy bin (and
correspondingly 100 events to get the same error on the flux in the
signal-limited regime) and the suppression of the systematic error on
the cross-over energy $E_{\mathrm{cross}}$. For the special case of
the pulsar observations, the cross-over energy can be significantly
reduced and will be close to $\sim25$ GeV (compared to $\sim40$ GeV in
the standard case of 5$\sigma$ and 10 events and 1\% systematic error
on the background flux). 

\section{Integral Sensitivity}

\begin{figure}[h]
\mbox{\epsfig{file=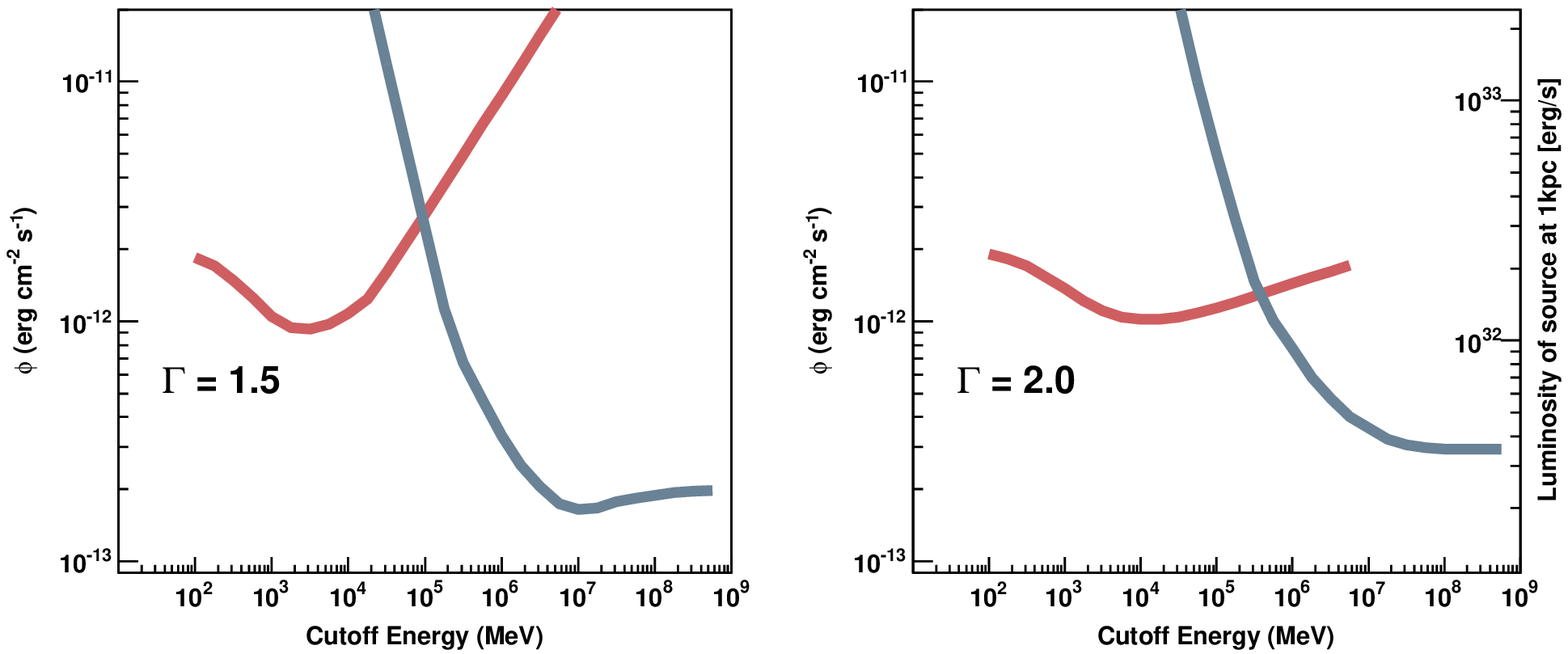, width=1.0\textwidth }}
\caption{Minimum detectable integrated (from 100 MeV to 100 TeV)
  energy flux, $\phi = \int{E dN/dE dE}$, as a function of cut-off
  energy $E_{c}$ for spectra of the form $dN/dE\propto
  E^{-\Gamma}\exp{-E/E_{c}}$.  Estimated CTA performance for 100 hours
  of on-axis observations of a point-like source (dark blue curves)
  are compared to a 10 year Fermi all-sky survey (red curves). The
  left hand plot shows the case of photon index $\Gamma$=1.5 and the
  right hand plot $\Gamma=2$. The equivalent luminosity for a source
  at 1~kpc distance is shown for reference. At least 10 detected
  photons, a $5\sigma$ detection and a signal to noise of better than
  1/20 are required in all cases.  }
\label{fig:4}
\end{figure}

The potential of an instrument to discover new sources, or events, is
related to its integrated performance over the relevant energy
range. Unfortunately estimates of integral flux sensitivity are always
strongly dependent on the assumed spectral shape. The most common
approach assumes a power-law spectrum of a given spectral index
($\Gamma$) and calculates the minimum flux ($F_{\mathrm{min}}$) above
a given energy ($E_{0}$) that is required for detection. As the
minimum flux in terms of photon rate per unit area is often a rapidly
falling function of energy $E\,\times\,F_{min}$, a quantity with the
same units as $\nu F_{\nu}$, is often plotted. Implicit in this method
is that the source spectrum starts abruptly at $E_{0}$ (or that all
information below $E_{0}$ is disregarded) and that the source spectral
power-law extends to infinity. Both of these assumptions are highly
unrealistic in practice. Here we adopt an alternative approach to
estimate minimum detectable flux, based on a characteristic energy in
the source spectrum: either a spectral energy distribution (SED) peak
or a cut-off energy. We define $\phi(E_{c})$ as the minimum detectable
integrated energy flux over the \emph{full energy range} of a source
with a spectrum $dN/dE\propto E^{-\Gamma}\exp{-E/E_{c}}$. A significance
of $5 \sigma$ and a minimum number of 10 events were requested (this
section deals mostly with the detection of sources, not so much with
the measurement of spectra).  Figure~\ref{fig:4} shows this
quantity for Fermi and CTA for two choices of $\Gamma$, 1.5 and
2.0. In the case $\Gamma=1.5$, $E_{c}$ corresponds roughly to the SED
peak energy. The minima in $\phi$ seen for Fermi at $\sim3$ GeV and
CTA at $\sim10$ TeV in the left panel of Figure~\ref{fig:4} can be
interpreted as the energies at which these instruments are most
sensitive for source detection. The cross-over between Fermi and CTA
is at a cutoff energy of $\sim90$ GeV. The case $\Gamma=2$ is also
interesting, demonstrating how dramatic the impact of a cut-off in the
source spectrum is on the detection probability: for CTA a cut-off at
100 GeV raises the minimum required source power by an order of
magnitude with respect to a $>1$ TeV cut-off and a detection of such a
source with Fermi is much more likely for the observation times
assumed (10 years for Fermi and 100 hours for CTA). The cross-over
between Fermi-LAT and CTA in this curve is at a cutoff energy of
$\sim370$ GeV. 

To help relate these curves to detection sensitivity for astrophysical
objects, the right-hand axis of Figure~\ref{fig:4} gives the
corresponding source luminosity at 1~kpc. For reference a canonical
proton accelerating SNR at this distance with an ambient density of 1
hydrogen atom per cubic centimetre would have a luminosity of
$\sim10^{50}$ erg/$t_{pp\rightarrow\gamma} \sim 2\times10^{34}$ erg/s,
suggesting that CTA should see such objects over most of the volume of
the Galaxy. As can be seen from this figure: for hard-spectrum
($\Gamma=2$) low-integrated luminosity sources such as SNRs, CTA will
ultimately perform better than Fermi unless the cutoff is below $\sim
1$~TeV.  We note that the case $\Gamma=1.5$ is close to the
expectation for dark matter annihilation spectra, (see
e.g.~\cite{Bergstroem1998, dm_paper}), or for a cut-off inverse
Compton spectrum from an uncooled electron spectrum (see
e.g.~\cite{HintonHofmann}). In the dark matter case the cut-off occurs
at a factor of up to a few (e.g.\ in the case of annihilation into
$b\bar{b}$) below the mass of the annihilating particle:
Figure~\ref{fig:4} therefore suggests that Fermi may be more effective
than CTA in searches for point-like dark matter annihilation
signatures for particle masses below $\sim300$ GeV.

\section{Sensitivity for Extended sources}
\label{sec:ext}
\begin{figure}[h]
\mbox{\epsfig{file=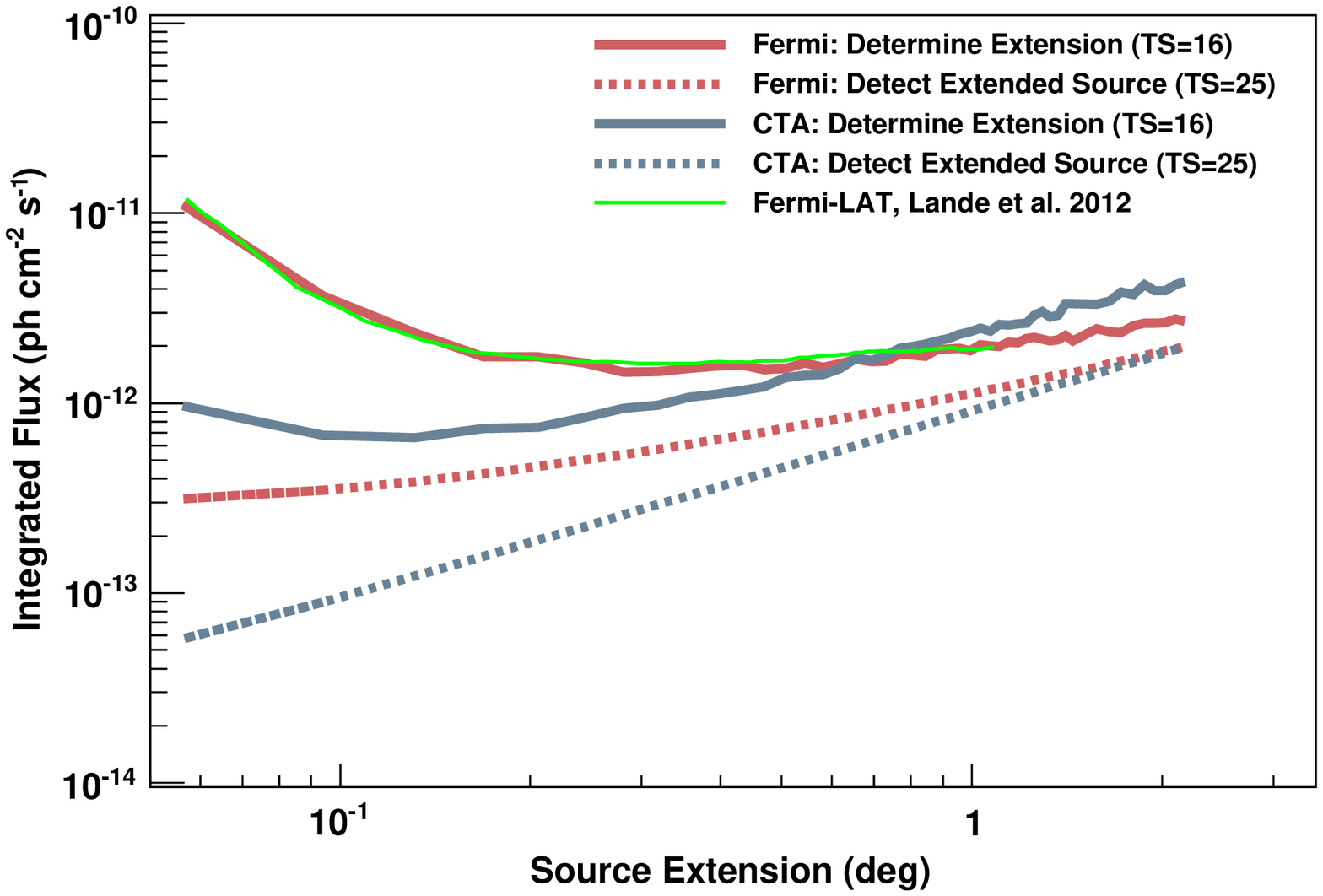, width=1.0\textwidth }}
\caption{Sensitivity for detecting extended sources as a function of
  the RMS (68\% containment) source extension. Shown is the integral
  flux $F_{d}$ at which an extended source is detectable at the $5
  \sigma$-level (dashed) and the integral flux at which extension can
  be demonstrated (solid) at the $4 \sigma$-level. See text for
  details of these quantities. Shown as a solid green line is the
  sensitivity estimate for extended sources given by the Fermi-LAT
  collaboration~\cite{FermiExtensionCat}. It can be seen that the
  detection sensitivity $F_{d}$ deteriorates more rapidly for CTA than
  for the Fermi-LAT due to the superior angular resolution. It can
  also be seen from the solid lines: up to sources of
  $\sim0.7^{\circ}$ CTA will perform significantly better in resolving
  sources.}
\label{fig:5}
\end{figure}

The power to detect regions of extended emission and to image/resolve
such regions is a key performance criterion for a gamma-ray
detector. A substantial fraction of the sources visible above 100 GeV
are significantly spatially extended, with a typically rms angular
size of $\sim0.2^{\circ}$ for Galactic objects
\cite{HESS:scanpaper1}. Whilst relatively fewer extended objects are
known at GeV energies, this may be, at least in part, a selection
effect \cite{FermiExtensionCat}. In addition to the dominant class of
extended Galactic objects, extended emission is expected from
extragalactic objects for some of the most important targets for CTA
and the Fermi-LAT, in particular for cosmic-ray and/or dark matter
annihilation signatures in clusters of galaxies and nearby
galaxies. For an extended object, at least 3 flux levels are of
interest: a minimum detectable flux $F_{d}$, the minimum flux $F_{e}$
at which statistically significant extension can be demonstrated and
the flux level $F_{i}$ at which substructure on the scale of the PSF
can be detected. Whilst $F_{i}$ can be readily estimated from the
point-source detection sensitivity $F_{ps}$: $F_{i} \approx F_{ps}
\Omega_{s}/\Omega_{psf}$ ($\Omega$ being the solid angle), the
remaining quantities are more subtle. An extended-source must be
detected above the fluctuations of an increased background level
$N_{bg,ext}(E) = N_bg(E) \sqrt{\theta_{psf}(E)^2 + \theta_{s}^2}$
where $\theta_s$ is the rms source size and $\theta_{\mathrm{psf}}$ is
the energy-dependent rms of the PSF, $F_d$ is therefore close to
$F_{ps}$ for $\theta_s<\theta_{psf}$. In contrast the ability to
detect the extension of a source improves dramatically for $\theta_s >
\theta_{psf}$.  Figure~\ref{fig:5} shows these two flux levels for
both Fermi and CTA for nominal observation lengths and assuming an
$E^{-2}$ spectrum source with no cut-off. In the Fermi case the
calculation of the Fermi collaboration for $F_{e}$ is shown in solid green for
comparison~\cite{FermiExtensionCat}. Figure~\ref{fig:5} shows that the
deterioration of source detection power for larger angular size
sources is much more dramatic for CTA than for Fermi, due to the
superior angular resolution of the former. Conversely, CTA can detect
modest source extension (at the $\sim0.1^{\circ}$ level) for order of
magnitude dimmer objects.  Note that the arguments given here apply
only to objects much smaller than the FoV of CTA (i.e. less than
$<<4^{\circ}$ rms) for which the on-axis response can legitimately be
assumed.


\section{Summary and Outlook}
This paper presents the first in-depth comparison of the sensitivities
in the overlapping energy range between 10~GeV and 100~GeV for the
Fermi-LAT and CTA. This is an important energy range due to the fact
that the Universe goes from being transparent to being opaque to
gamma-rays at these energies. It also is the range at which many
Fermi-LAT sources show interesting features in their spectra, such as
cutoffs and breaks or new components. When comparing the differential
sensitivity of a 10-year sky-survey with the Fermi-LAT with a typical
100 hour exposure of CTA we find that CTA will be better for measuring
spectra (taking into account systematic errors)  for an $E^{-2}$
source above $\sim40$ GeV. In terms of detecting sources, CTA will
work better for sources with an $E^{-2}$ spectrum and and an
exponential cutoff above $\sim370$ GeV. For short-term phenomena
(order of minutes) CTA will perform orders of magnitude better than
the Fermi-LAT in the overlapping area, although the Fermi-LAT
obviously has a huge advantage in terms of field of view.
Given the large overlap in energy range, an ideal scenario is one
where both the Fermi-LAT and CTA operate simultaneously during some
(albeit brief) period of time. In fact, CTA would benefit tremendously
from a simultaneous operation with the Fermi-LAT. Therefore, the
ground-based gamma-ray community is strongly in favor of extending the
lifetime of the Fermi-LAT mission over the currently planned 5-7
years.

\section*{Acknowledgements}
We gratefully acknowledge support from the following agencies and
organisations: Ministerio de Ciencia, Tecnolog\'ia e Innovaci\'on
Productiva (MinCyT), Comisi\'on Nacional de Energ\'ia At\'omica (CNEA)
and Consejo Nacional de Investigaciones Cient\'ificas y T\'ecnicas
(CONICET) Argentina; State Committee of Science of Armenia; Ministry
for Research, CNRS-INSU and CNRS-IN2P3, Irfu-CEA, ANR, France; Max
Planck Society, BMBF, DESY, Helmholtz Association, Germany; MIUR,
Italy; Netherlands Research School for Astronomy (NOVA), Netherlands
Organization for Scientific Research (NWO); Ministry of Science and
Higher Education and the National Centre for Research and Development,
Poland; MICINN support through the National R+D+I, CDTI funding plans
and the CPAN and MultiDark Consolider-Ingenio 2010 programme, Spain;
Swedish Research Council, Royal Swedish Academy of Sciences financed,
Sweden; Swiss National Science Foundation (SNSF), Switzerland;
Leverhulme Trust, Royal Society, Science and Technologies Facilities
Council, Durham University, UK; National Science Foundation,
Department of Energy, Argonne National Laboratory, SLAC National
Accelerator Laboratory, University of California, University of
Chicago, Iowa State University, Institute for Nuclear and Particle
Astrophysics (INPAC-MRPI program), Washington University McDonnell
Center for the Space Sciences, USA. The research leading to these
results has received funding from the European Union's Seventh
Framework Programme ([FP7/2007-2013] [FP7/2007-2011]) under grant
agreement no 262053.

\bibliographystyle{elsarticle-num}
\bibliography{ctafermi}

\end{document}